# Evaluating Electronic Customer Relationship Management Performance: Case Studies from Persian Automotive and Computer Industry

**Narges Safari**
School of Business Information Systems
Auckland University of Technology
Auckland, New Zealand
Email: Narges.Safari@aut.ac.nz

**Fariba Safari**
School of Computer Science
Eindhoven University of Technology
Eindhoven, Netherlands
Email: f.safari@tue.nl

**Karin Olesen**
School of Business information Systems
Auckland University of Technology
Auckland, New Zealand
Email: karin.olesen@aut.ac.nz

**Fatemeh Shahmehr**
School of Management
Monash University
Melbourne, Australia
Email: shahmehr.f@gmail.com

## Abstract

This research investigates the influence of industry on electronic customer relationship management (ecrm) performance. A case study approach with two cases is applied to evaluate the influence of ecrm on customer behavioral and attitudinal loyalty along with customer pyramid. The cases covered two industries consisting of computer and automotive industries. For investigating customer behavioral loyalty and customer pyramid companies' database are computed while for examining customer attitudinal loyalty a survey is conducted. The results show that ecrm has significantly different impacts on customer behavioral and attitudinal loyalty and customer pyramid in two industries. These results indicate that organisations need to consider their specific industry and company requirements before establishing their ecrm systems. Moreover, this research shows that how organisations can evaluate their ecrm performance based on customer attitudinal and behavioral loyalty and customer pyramid and formulate new policies to improve relationship strategies with customers.

**Keywords**

Electronic Customer relationship Management, Customer Behavioral Loyalty, Customer Attitudinal Loyalty, Customer Pyramid.

## 1   Introduction

An increasing number of organizations are applying electronic customer relationship management (ecrm) systems to attract and retain valuable customers and to improve analytical capabilities (Fjermestad and Romano, 2003). In fact ecrm is a part of a complete customer relationship management (crm) strategy (Feinberg et al., 2002). Incorporating people, process, and information systems, crm strategy produces seamless integration among every part of business that have interaction with customers. While, with the arrival of internet-based technologies, ecrm develop the traditional crm by integrating people, process, and information systems through internet-based channels (Kim et al., 2002). In comparison with crm, ecrm provides new opportunities for companies to save more money. For instance, Sun Microsystem saved 1.3$ million per year after applying ecrm (Feinberg et al., 2002). Many companies are interested in improving relationships with customers

through ecrm systems. This is apparent for companies in the finance (Zuccaro and Savard 2010) and manufacturing industries (Sivaraks et al., 2011). Therefore, several studies highlight ecrm advantages for different types of organisations. For instance, Lee-Kelley et al., (2003) finds that ecrm provides the promise of increased customer loyalty for e-retail companies. Ecrm improves the number of frequent purchases and customer behaviours, and intentions in Malaysian IT help desk industry (Abdullateef et al., 2014, Ghandehary et al., 2014).

However, in some cases ecrm project was not successful as expected. For example, Hadaya and Cassivi (2009) demonstrates that ecrm cannot improve the relationships between interorganisational process and key customers in IT service providers, resulting in decreasing customer loyalty which is one of the main purpose of implementing ecrm. Wang et al., (2014) states that the main reason for ecrm failure is that companies are weak in measuring the outcomes of ecrm systems accurately. It means that, when companies examine the success of their ecrm regularly, they will be able to understand its weaknesses and formulate new policies to improve them. Therefore, in this study, we examines the success of ecrm in two different industries: computer and automotive industries to see if ecrm performance significantly differs in two industries. Marketing strategies has shifted to customer retention and customer loyalty. The main aim of ecrm is improving customer loyalty and motivating valuable customers to remain loyal (Fjermestad and Romano 2003). Thus, in this study, we evaluate customer loyalty (attitudinal and behavioural loyalty) and pyramid of each company and compare the results to see in which company ecrm is more successful in having more loyal and more valuable customers in the top layer of customer pyramid.

Customer loyalty is defined as the commitment and positive attitudes of customers toward a company, which leads to repurchasing its products in the future (Khan 2013). In this definition, there are two key terms that need to be considered in analysing customer loyalty: 1- commitment and positive intention which refers to customer attitudinal loyalty and 2- repurchasing companies' products in the future which refers to customer behavioural loyalty. In this article, we examine both customer attitudinal and behavioural loyalty along with customer pyramid. Customer pyramid is a successful method for differentiation of customers determining valuable customers. Customer pyramid categorise customers into meaningful groups by which companies can determine their valuable customers (Wei et al., 2010). The rest of this paper is organised as follows: in section 2, through a critical review of literature, we define a series of research hypotheses. Section 3 explains the research methodology. Data analysis techniques and findings are described in section 4, section 5 explains discussion, research limitations, future research directions, and managerial implications. Finally, section 6 provides a brief conclusion.

## 2  Literature Review

Customer loyalty is defined as the commitment and positive attitudes of customers toward a company, which leads to repurchasing its products in the future (Khan 2013). In this definition, there are two key terms that need to be considered in analysing customer loyalty: 1- commitment and positive intention which refers to customer attitudinal loyalty and 2- repurchasing companies' products in the future which refers to customer behavioural loyalty. Both attitudinal and behavioural loyalties are important for companies. It means that customers with high level of loyalty not only have emotional commitment to a company, but also make more frequent purchases from that company (Doherty and Nelson, 2008).

Several studies have examined the attitudinal and behavioural loyalty in different industries. Shang and Lin (2010) confirm the positive link between ecrm and customer behavioural loyalty in car dealership sector and telecommunications industry. While, applying survey data, Del Castillo Peces et al., (2012) reports that through implementing ecrm, 50% of Spanish banks have not received the significant level of customer behavioural loyalty. They mention that these banks need specific policies in order to improving their customer behavioural loyalty through ecrm. Moreover, analysing customer attitudinal loyalty, Shim et al., (2012) categorise customers, of an online-shopping mall, into VIP and non-VIP groups and state that this mall should formulate specific policies in interacting with each group. Khalifa and Shen (2005) prove that ecrm has a positive effects on customer attitudinal loyalty in the hardware retailing industry. Thus, we can draw conclusions that ecrm directly influences customer loyalty: behavioural and attitudinal loyalty. However, its performance seems to be different from industry to industry:

H1: the ecrm success based on customer attitudinal loyalty in computer industry significantly differs from automotive industry.

H2: the ecrm success based on customer behavioural loyalty in computer industry significantly differs from automotive industry.

The most popular and successful method for selection and differentiation of customers is customer pyramid theory (Wei et al., 2010). In this method customers are partitioned into analogous groups (Farzipoor Saen 2013) so that each group can be addressed in a unique marketing strategy (Buttle 2009). The customer pyramid enables companies to raise customers' profitability by improving the link between service quality and profitability. According to Zeithaml et al., (2001) and Wei, Lin et al., (2010), the customer pyramid is divided into four categories: (1) Platinum Level (these customers are very profitable and loyal). (2) Gold Level (these customers have less profitability than platinum level). (3) Iron Level (these customers possess loyalty and profitability levels that are not enough to receive special services). (4) Lead Level (These customers are costly and have low profitability). Different studies apply customer pyramid to evaluate the effectiveness of the relationship strategies between companies and their customers. For example, Zeithaml et al., (2001) applies customer pyramid with four levels to examine the success of relationship strategies between a service provider and its customers and to recognise the most profitable and the least profitable customers. Curry (1997) builds customer pyramid for several companies in different industries, including electronic component manufacturers, wholesalers, automotive dealers, retailers, and banks, to examine their customer behaviour. Interestingly, he finds that customer pyramids of all companies have notably similar customer behaviour patterns, which shows that their relationship strategies with customers are similar. Customer relationship strategies directly influences the customer pyramid. Building customer pyramid for an automotive company, Zeithaml et al., (2001) shows that ecrm system moved 20% of lead customers to the Iron category and 10% of the Iron customers to the Gold category. In the current study, we aim to compare the success of ecrm strategy between computer industry and automotive industry. Therefore:

H3: the ecrm success based on the rate of customer pyramid in Computer Industry significantly differs from automotive industries.

## 3　Research Methodology

This paper aims to compare the ecrm performance in two industries: computer industry and automotive industry. This paper is a cross-sectional study using two different cases (Sahai and Khurshid 1995). Unlike case-control studies, the cross-sectional study as a form of epidemiological study, can be used to gather targeted data from different population at one specific point in time (Breslow and Day 1987). As the participants in this study are customers and the results are collected from two industries, a cross-sectional study is appropriate to this research. As shown in Figure 1 and 2, we analyse costumer attitudinal loyalty for each company and compare the results to see if the attitudinal loyalty of computer industry is significantly different from automotive industry. For analysing attitudinal loyalty we apply survey data. For analysing attitudinal loyalty survey is an appropriate tool to measure customers' commitment to company (Del Castillo Peces et al., 2012).

In the next step, we evaluate the customer behavioural loyalty of each company and compare the result to see if the customer attitudinal loyalty of computer industry is different from automotive industry. In order to analysing customer behavioural loyalty RFM method is applied. RFM is an acronym of three behavioural variables: Recency of purchases (time elapsed since last purchase or R), frequency of purchases (number of purchases in a given time period or F) and monetary value of purchases (monetary value of purchases in a given time period or M) (Ha and Lee 2010). This technique is mostly applied for evaluating customers' behavioural patterns and recognising loyal customers (Liao et al., 2014). For calculating RFM value, R, F, and M are combined (Lejeune 2001). We collect R, F, and M data from companies' databases.

In the last step we build customer pyramid for each company and compare the number of customers of each segment to see if there is a significant difference between the number of valuable customers of company industry and automotive industry. In order to building customer pyramid, an appropriate customer segmentation technique should be applied. Among all clustering techniques, RFM has recently received further attention in various studies aiming to evaluate ecrm performance (Coussement et al., 2014). Indeed, the RFM technique has been well-established as a successful technique in customer segmentation and analysis (Chang and Tsai 2011). Figure 1 and 2 explain our research methods. Figure 1 describes how we collect our survey data and analyse customer attitudinal loyalty. While, Figure 2 presents the analytical steps to evaluate customer behavioural loyalty and pyramid. As shown in Figure 2, we use RFM method for two purposes: analysing customer behavioural loyalty and categorising customers in customer pyramid.

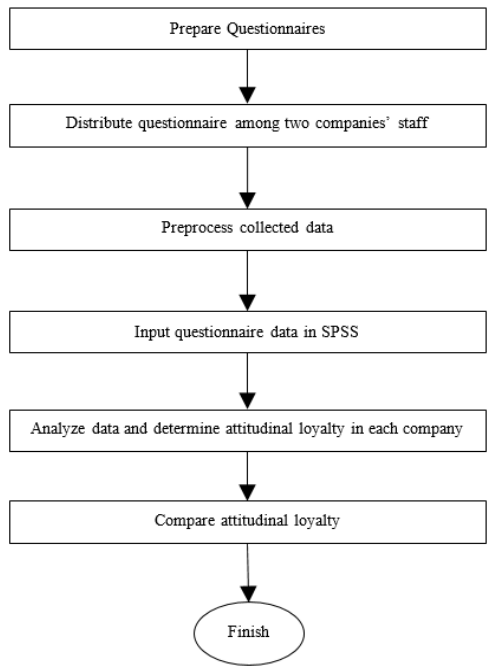

*Figure 1 Customer attitudinal loyalty*

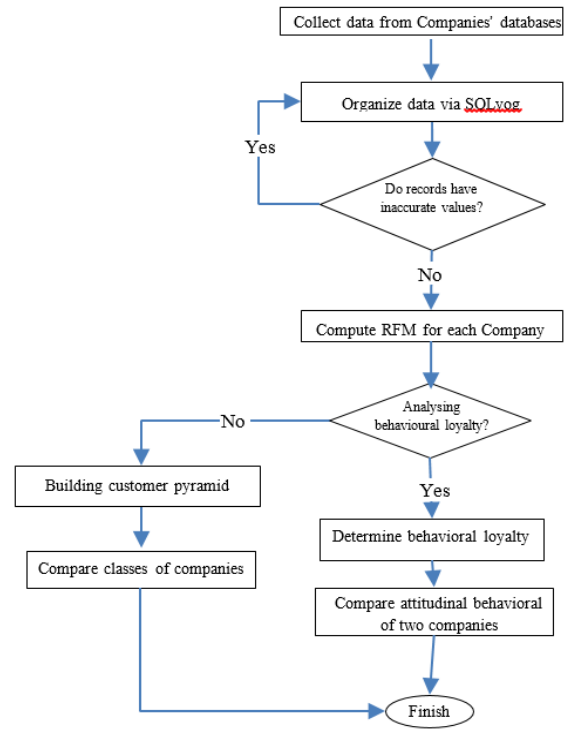

*Figure 2 Customer behavioral loyalty*

### 3.1 Case Studies' Backgrounds

We collect our data from two industrial companies: computer and automotive companies. We select these companies because: they are using ecrm system since 2008 and they are the biggest and most well-known companies in their industries in Iran.

*Computer Company:*

The mission of this company is providing high-quality optical carriers of information under the trade mark. This company produces 60 different Pre-recorded Media (CD-ROM, DVD-ROM) brands. It is the largest producer of blank and pre-recorded CD and DVD in Middle East and the 4th largest factory between 118 other factories in the world. The company exports its products to European and Far Eastern countries such as Poland, Slovenia, Turkey, Italy, Singapore, and India. The main products of this company are CD/DVD, games along with mobile and Tablet accessories.

*Automotive Company:*

The mission of automotive Company is the design, production, and sale of heavy and light, taxi and medical vehicles. This company produces 9 different vehicle brands. It is the largest vehicle manufacturing company in Iran. It has 13 overseas branches consisting of CIS, Middle East, North Africa and East Europe regions.

## 4 Data Analysis

In this section, we explain data analysis steps for evaluating attitudinal loyalty, behavioral loyalty, and customer pyramid separately.

### 4.1 Customer Attitudinal Loyalty

For evaluating customer attitudinal loyalty, as shown in Figure 1, we collect data through questionnaire with 5 items on the 5 points Likert's scale. In order to measuring the content validity of the questionnaire, it was distributed among 15 experts in the field of ecrm, consisting alumni and faculty members of various universities in Iran. Experts are often assumed to be professionally qualified and/or to have achieved great status (Mullen 2003). In this study, an expert is one instructed by experience; a person is sufficiently qualified in any special branch of learning including ecrm and customer-organization relationship. For examining experts 'opinions, the binomial test was employed (sign-level = 0.05, cut-point = 3) by SPSS software. According to Table 1 the significant levels for all indices are lower than 0.05 and the frequency of observations for the group 1 (>3) are greater than group 2, so the measurements are approved with 95% confidence.

| **Items** of attitudinal loyalty | Segments | Category | N | Observed prop. (%) | Exact sig. (2-tailed) | Reject/ confirm |
|---|---|---|---|---|---|---|
| Commitment to company | 1 | ≤ 3 | 1 | 0.06 | 0.001 | Confirm |
|  | 2 | >3 | 15 | 0.94 |  |  |
|  | Total |  | 16 | 1.00 |  |  |
| Close relationship with company | 1 | ≤ 3 | 1 | 0.06 | 0.001 | Confirm |
|  | 2 | >3 | 15 | 0.94 |  |  |
|  | Total |  | 16 | 1.00 |  |  |
| The first priority for shopping is this company | 1 | ≤ 3 | 0 | 0.00 | 0.001 | Confirm |
|  | 2 | >3 | 16 | 1.00 |  |  |
|  | Total |  | 16 | 1.00 |  |  |
| Never abandon this company | 1 | ≤ 3 | 1 | 0.06 | 0.001 | Confirm |
|  | 2 | >3 | 15 | 0.94 |  |  |
|  | Total |  | 16 | 1.00 |  |  |
| Never change my beliefs about company | 1 | ≤ 3 | 1 | 0.06 | 0.001 | Confirm |
|  | 2 | >3 | 15 | 0.94 |  |  |
|  | Total |  | 16 | 1.00 |  |  |

*Table 1 the results of the binomial test (sig. level=95%)*

### 4.1.1 Data Collection

Data were collected from companies' customers purchasing products through companies' websites. The population of customers of computer company was 235 customers and the population for automotive company was 178 per day. We uploaded online questionnaire in both companies' website and collected 160 completed questionnaire in each company. Based on Krejcie and Morgan's table, this sample size (160) is reasonable.

### 4.1.2 Data analysis

To examine reliability of the questionnaire, Cronbach's alpha coefficient was applied. Calculated Cronbach's alpha coefficient is 0.854 in the expert's survey. It is higher than 0.70 that proves the questionnaire is reliable (Hinton et al., 2014). For evaluating customers' attitudinal loyalty, Mann–Whitney U test was done. As shown in Table 2, the value of |Z| is 11.650 that is higher than 1.96 and significance level is 0.00 that is lower than 0.01 (Rosner and Grove 1999). Therefore, H1 is confirmed with 99% confidence.

|  | **Q1-Q5 average** |
|---|---|
| Mann-Whitney U | 2556.5 |
| Wilcoxon W | 13881.5 |
| Z | -11.65 |
| Asmp. Sig. (2-tailed) | .000 |

*Table 2 the result of hypotheses testing*

According to Table 2, the ecrm success, based on customer attitudinal loyalty, in computer industry significantly differs from automotive industry.

## 4.2 Customer Behavioural Loyalty

According to literature of review, RFM model was used to measure customer's behavioral loyalty.

### 4.2.1 Data Collection

For evaluating customer behavioural loyalty we applied RFM method. These customers' records (Recency, Frequency, and Monetary) were collected from companies' databases. Monetary and Frequency are numeric values, while the type of Recency is a date. Thus, to extract its numeric value, Structural Query Language (SQL) queries in SQLyog software were used. SQL is a programming language designed for data management of database. A function for converting date to number is written in which the interval between the date of the most recent purchase and the analysing time was calculated (Wei et al., 2010).

### 4.2.2 Data Analysis

In order to score customers' RFM values, one scoring method is customer quintile method in which customers' database will be divided into five equal quintile. This method of RFM scoring has some challenges, as dividing customers into equal quintile would segment customers with identical behavior into different groups. Therefore, in this study another scoring method called behavior quintile scoring method is used in which RFM values are divided into five groups with different number of customers. The top quintile is coded as 5 and the next is coded as 4 and so on (Miglautsch 2000, Wei et al., 2010). The boundaries are determined based on sales expert's opinions. Since, both companies follow the same sale rules, so they have equal ranking (see Table 3).

| Monetary (Rial) | Frequency | Recency | Rank | Code |
|---|---|---|---|---|
| $M \geq 750{,}000{,}000$ | $F \geq 50$ | $R \leq 7$ | Very High (VH) | 5 |
| $750{,}000{,}000 > M \geq 500{,}000{,}000$ | $30 \leq F < 50$ | $7 < R \leq 14$ | High(H) | 4 |
| $500{,}000{,}000 > M \geq 250{,}000{,}000$ | $15 \leq F < 30$ | $14 < R \leq 30$ | Medium (M) | 3 |
| $250{,}000{,}000 > M \geq 50{,}000{,}000$ | $5 \leq F < 15$ | $30 < R \leq 90$ | Low(L) | 2 |
| $M < 50{,}000{,}000$ | $F < 5$ | $R > 90$ | Very Low(VL) | 1 |

*Table 3 RFM scoring rules for both companies*

To extract R, F, M codes from database, Nested IF in Excel has been used, for instance, the following formula is used to calculate M values (formula 1).

$$= \mathrm{IF}([@Monetary] \geq 750{,}000{,}000, 5, \mathrm{IF}([@Monetary] \geq 500000000, 4, \mathrm{IF}([@Monetary] >= 250{,}000{,}000, 3, \mathrm{IF}([@Monetary] >= 50{,}000{,}000, 2, 1))))$$

(Formula 1)

To compare RFM mean in two companies, the acquired data is analyzed by student independent t-test in SPSS16 software with significant level of 95%. Since the sig level is 0.000 and is less than 0.05, H2 is admitted. This proves that ecrm is more effective on automotive company in comparison with computer company.

## 4.3 Customer Pyramid

In order to building customer pyramid for each company, RFM model was applied. Several studies have categorised customers based on RFM model. For instance, applying RFM model, Pitta et al., 2006 built customer pyramid for one bank and proposed that this bank should apply following relationship strategies regarding each pyramid level: this bank should offer a high level of services to their platinum and gold customers. It should have a respectful manner with their Gold clients while it withheld customer service strategy from Lead customers. For clustering customers based on RFM model, we need a clustering algorithm. K-means is one of the simplest and most popular algorithms for clustering and used in various fields including data mining, statistical data analysis and other business applications. Applying K-means method through RFM, Liao et al., (2014), segment customers of the hairdressing industry into four sections and developed new strategies for each category.

In this study we apply K-mean clustering algorithm for categorising customers based on RFM model. The outputs of the RFM method became the inputs of the k-means algorithm for the customer pyramid. Clustering is the process of extracting homogenous groups from a set of data. K-means algorithm is a well-known clustering algorithms that partition a set of data into different groups (Jain 2010). Cluster analysis techniques consist of hierarchical and partitioning algorithm. The former method provides data as a hierarchical decomposition, while the latter one creates cluster in which data points are highly similar. K partitions are defined in partitioning algorithms (Kalyani and Swarup 2011). RFM and K-means algorithm were employed to cluster customer value of a company of Taiwan's electronic industry (Cheng and Chen 2009) and Taiwan automobile dealer (Tsai et al., 2013). Hosseini et al., (2010) also used this algorithm for customer segmentation in automotive industry, they have confirmed that by expanding RFM into K-means algorithm, an efficient customer classification helps organizations to improve their ecrm strategies. In the current study, this algorithm is used for customer segmentation in two companies.

### 4.3.1 Data Analysis

In this step we used RFM records of companies' customers. K-means clustering algorithm in MATLAB R 2011, applied to cluster RFM records into four sections of customer pyramid. Output of k-means algorithm is shown in Table 4 for computer company and in Table 5 for automotive company wherein number of customers, percent of customers, and customer pyramid class for each cluster is achieved based on R, F and M values. Figure 3 illustrates the percent of customers for classes of both companies (platinum, gold, iron, lead).

| Cluster | R | F | M | RFM | Number of customers | Percent | Class |
|---|---|---|---|---|---|---|---|
| 1 | 4.59604 | 2.394059 | 1.994059 | 8.984158 | 505 | 31.5625 | Iron |
| 2 | 2.487091 | 2.056799 | 2.003442 | 6.547332 | 581 | 36.3125 | Lead |
| 3 | 3.421384 | 2.22327 | 4.820755 | 10.46541 | 318 | 19.875 | Gold |
| 4 | 3.770408 | 4.484694 | 2.591837 | 10.84694 | 196 | 12.25 | Platinum |

*Table 4 K-means cluster results for computer company*

| Cluster | R | F | M | RFM | Number of customers | Percent | Class |
|---|---|---|---|---|---|---|---|
| 1 | 4.658182 | 3.09697 | 1.452121 | 9.207273 | 825 | 17.92699 | Iron |
| 2 | 2.395215 | 1.353135 | 1.223597 | 4.971947 | 1212 | 26.33638 | Lead |
| 3 | 4.295266 | 4.969231 | 4.881065 | 14.14556 | 1690 | 36.58316 | Platinum |
| 4 | 3.393143 | 2.156571 | 5 | 10.54971 | 875 | 19.01347 | Gold |

*Table 5 K-means cluster results for automotive company*

As it can be seen in Figure 3, the percent of customers in Platinum class for automotive company (%36.58), is greater than computer company (%12.25); but the percent of customers in Iron and Lead classes for automotive company (%17.93, %26.37) is less than the same ratios for automotive company, by %35.56 and %36.31 respectively. There is no significant difference between the same ratios of Gold class for the two companies. It is obvious that the ratio of customers in both classes of Gold and Platinum indicates the success of company in relationship management with them. In another words, if the number of customers in these two classes were greater than the same number of the other two ones (Lead and Iron), we can conclude that the tool that has been used (ecrm), were more customers' center, as it maintained more loyal customers.

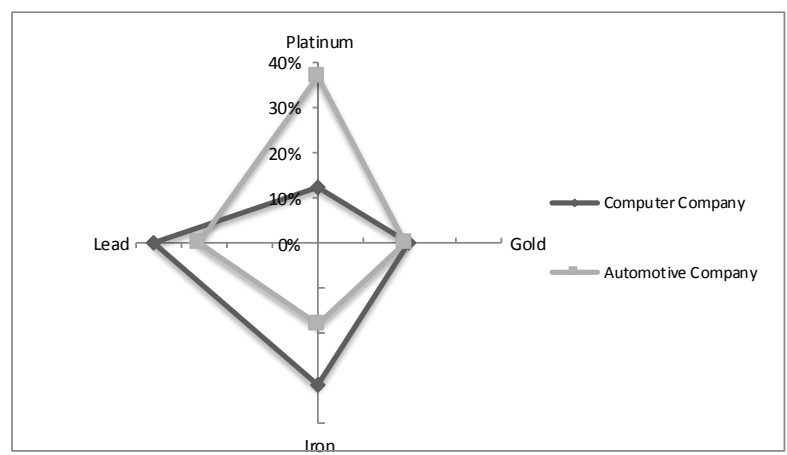

*Figure 3 Percent of customers in each class*

According to Table 6, the percent of customers in Platinum and Gold classes for automotive company (%55.73) is greater than the percent of customers in the same classes for computer company (% 32.13). However, for the other two classes, the case is different; the percent of customers in Lead and Iron classes for automotive company (%44.27) is less than the percent of customers in the same classes for computer company (% 67.87).

| Classes | Computer Company | Automotive Company |
|---|---|---|
| The percent of customers in gold and platinum levels | 32.13 | 55.73 |

*Table 6 Comparison between the percentages of companies' customers in the four classes*

In order to investigate H3, success proportion was examined. In this case gold and platinum customers are categorized as the success group for each company. Table 7 shows the results (The sample success proportion or observed proportion of successes in a sample is shown with $\bar{p}$ and the population success proportion is shown with *P*).

| company | gold | Platinum | X | n | The sample success proportion |
|---|---|---|---|---|---|
| Computer | 318 | 196 | 514 | 1600 | $\bar{p1}$=0.321 |
| Automobile | 875 | 1690 | 2565 | 4620 | $\bar{p2}$=0.555 |

*Table 7 the success proportion*

As is shown in Table 7 $\bar{p2} > \bar{p1}$. We also consider the use of the observed success proportion $\bar{p}$ to evaluate the value of the population success proportion *p*:

$$\bar{p} = \frac{X1+X2}{n1+n2} = 0.495$$

X1= 514

X2= 2565

n1= 1600

n2=4620

$\bar{p1}$=0.321

$\bar{p2}$=0.555

$$Z = \frac{\bar{p1}-\bar{p2}}{\sqrt{\bar{p}(1-\bar{p})(\frac{1}{n1}+\frac{1}{n2})}}$$

Z=-16.13

Because Z is less than -1.96 so H3 is accepted with 95% significance. Therefore, we can conclude that automotive company has had a better combination of customers (more loyal customers) in comparison with computer company.

## 5 Discussion

This research contributes to marketing literature in which the impact of ecrm on behavioural and attitudinal loyalty along with customer pyramid are considered based on the type of industry. This research examined two databases from two major industries (computer industry and automotive industry). Hypothesis 1 evaluated whether or not the ecrm success based on customer attitudinal loyalty in computer industry differs from automotive industry. Findings show that ecrm leads to higher attitudinal loyalty in automotive industry compared to computer industry. Comparing this finding to a study by Shang and Lin (2010) that emphasizes positive attitudes toward frequent purchases in two industries (car-dealership and telecom industries), we found hypothesis 1 was not only confirmed but also supported findings in previous studies. This paper suggests that the lower behavioral and attitudinal loyalty in computer industry may owe to a lack of well-matched ecrm policies with the features of computer industry. As suggested by Steel et al., (2013), ecrm design and its implementation methods should be selected based on the industry context. A potential implication of confirmation of both hypotheses is that they support the importance of 'industry' as a variable in creating ecrm polices.

Hypothesis 2 investigated whether or not the ecrm success based on customer behavioural loyalty in computer industry differed from automotive industry. Findings indicate that ecrm leads to higher behavioural loyalty in automotive industry in comparison to computer industry. This finding is similar to a study by Lee-Kelley et al., (2003) suggesting that customers have behavioural loyalty to e-retailer companies by frequent purchases, while a study by Sophonthummapharn (2009) shows that the tendency of ecrm success in manufacturing enterprises relies more on recency in considering that customers had long-term purchases. Our findings support these differences by confirming the second hypothesis.

As ecrm success tends to be affected by marketing segmentation that improves the quality of services to various groups of customers based on their different attitudes and behaviours. In the next stage, this research categorized customers to four classes for each company. As a result, hypothesis 3 investigated whether or not the ecrm success based on class customer in customer pyramid in computer industry significantly differs from automotive industry. The results indicate that ecrm success in maintaining customers loyal in automotive company is greater than computer industry. With similar to Zeithaml et al., (2001) and Wei et al., (2010), this research highlighted the importance of market segmentation in e-CRM and confirmed the last hypothesis. Depending on what class the customer is, the ecrm success tends to be associated with the policies provided in various classes. For example, ecrm policy for a customer who is served in computer industry and located at Platinum class, differs from someone else who is served in automotive industry and located at Platinum class. Both companies must consider that these customers might not be interested in the same customer services. Unlike the study by (Cheng and Chen 2009), this research contributed to the influence of industry context on the success of ecrm.

## 5.1 Limitations and Further Research Directions

The research has some limitations. While cross-sectional case study is a useful methodology for evaluating two or more independent populations, more than two populations can provide better results. Moreover, this research only concentrates on computer industry and automotive industry. However, the findings might be different in other industries such as electricity/water supply, telecommunications, or insurance. We recommend that further research be conducted to investigate ecrm performance in other industries and compare their results with this research's findings. Moreover, in this study we evaluate ecrm success based on three factors (customer attitudinal and behavioural loyalty and customer pyramid). Future researchers can evaluate ecrm performance based on more factors like customer satisfaction and service quality.

## 5.2 Managerial Implication and Customer Benefits

Research on ecrm rarely emphasizes the managerial implications of industry type in creating and capturing new opportunities to maintain customers. The investigation on the influence of industry type on ecrm performance can have four managerial implications.

First, managers in companies that are interested in applying ecrm in practice can fully understand how to establish an ecrm based on specific company knowledge. For example, in an automotive industry, managers deal with a long-term customer support in which customer expect that companies provide them with a wide variety of service, including information services, safety services and repair and maintenance services. However, in the insurance industry, although customers are served on an ongoing basis, they might be more sensitive than the customers in the automotive industry because providing insurance services tends to be easier than after-sale services for a customer that has a vehicle. Therefore, mapping ecrm in a specific industry needs to consider the type and nature of service or product as well as the impact of customer service on operations.

Second, the results indicated that ecrm in a specific industry can differently affect customer behavioural and attitudinal loyalty. These findings encourage managers to create specific policies and strategies for maintaining customer interest in their products and services. For example, in e-publishing or software industry, customers prefer to contact customer service through online or over the phone (without face-to-face communications) to access services quickly. For these customers, behavioural loyalty can result from ease of access and the use and speed of service. Attitudinal loyalty can be seen in customer commitment to the company over an extended period of time. This loyalty can be fostered from building a positive attitude to products and services based on the customer behavioural loyalty. However, in the banking industry, customers seem to be more interested in financial transaction security. These customers prefer to have a specific customer support agent who they can trust. For these customers, an ecrm policy that addresses behavioural loyalty is likely to encourage new ways of approaching financial transactions, and attitudinal loyalty can be the result of 100% safety of transactions over the long-term, in turn increasing commitment to companies in this industry.

Third, managers can group their customers based on the industry that they have been working in. The customer pyramid enables companies to motivate customers in two ways: (1) Customers feel that they are very important to the company; (2) Customers can earn profitability from their commitment and the rate of their loyalty. However, the results of this clustering seem to vary from one industry to another. Therefore, when the percentage of customers belonging to a particular cluster is identified, managers can provide customers with targeted services and products. This percentage, in turn, can help them to share customer company profitability that can be appeared in the form of especial gifts, discounts and offs. In addition, managers can find new policies and strategies to grow the number of customers in the customer pyramid. For example, in this paper, the number of customers in Platinum class in automotive industry is higher than the number of customers in computer industry. This study has found that managers in automotive industry must make special offers to these customers, while the number of customers in Iron class and Lead class in computer industry is greater than the number of customers in automotive industry.

Overall, when an ecrm performs in an efficient way and is completely suited to a specific industry, customers are likely to be more satisfied and companies can help them to save money and time. An accurate ecrm can also facilitate the way customers interact with companies based on the type of service, its importance, and the class of customer in the customer pyramid. In addition, while an organization completely changes the way to interact with customers (e.g. from face-to-face to online interactions), it might lead to a decrease in customers interested in staying in touch with the company. An accurate evaluation of the suitability of an ecrm based, therefore, on the type of industry in reference to customer loyalty and the customer pyramid might give the company this opportunity to better serve customers.

# 6  Conclusion

This paper compared ecrm performance in two industries: computer and automotive industry. Ecrm performance was evaluated based on customer attitudinal and behavioural loyalty and customer pyramid. The results showed that ecrm has significantly different impacts on customer behavioural and attitudinal loyalty and customer pyramid in two industries. These results indicated that organisations need to consider their specific industry and company requirements before establishing their ecrm systems.